\documentclass[aps,prb,twocolumn,amsmath,amssymb,superscriptaddress]{revtex4}
\usepackage{bm}
\usepackage{graphicx}

\begin{document}

\title{Fluxon cotunneling in coupled Josephson junctions: Perturbation theory}
\author{Andrew G. Semenov}
\affiliation{I.E. Tamm Department of Theoretical Physics, P.N. Lebedev Physical Institute, 119991 Moscow, Russia}
\affiliation{Skolkovo Institute of Science and Technology, 121205 Moscow, Russia}
\author{Alex Latyshev}
\affiliation{Departement de Physique Theorique, Universite de Geneve, CH-1211 Geneve 4, Switzerland}
\author{Andrei D. Zaikin}
\affiliation{I.E. Tamm Department of Theoretical Physics, P.N. Lebedev Physical Institute, 119991 Moscow, Russia}
\affiliation{National Research University Higher School of Economics, 101000 Moscow, Russia}
\date{\today}
\begin{abstract}
We investigate the effects of fluxon cotunneling and quantum Coulomb drag in a system of two small Josephson junctions coupled 
by means of mutual capacitance $C_m$. Depending on the value of $C_m$ we identify three different regimes of strong, intermediate and weak coupling. Focusing our attention on the last two regimes we develop a perturbation theory in the interaction and explicitly derive
fluxon cotunneling amplitudes at sufficiently small mutual capacitance values. We demonstrate that the Coulomb drag effect survives at
any non-zero $C_m$ and evaluate the non-local voltage response that is in general determined by a trade off between two different cotunneling processes. Our predictions can be straightforwardly generalized to bilinear Josephson chains and directly verified in future experiments.

\end{abstract}
\maketitle

\section{Introduction}
\label{intro}

Macroscopic quantum phenomena in superconducting circuits involving ultrasmall Josephson junctions \cite{SZ90,book} attracted a lot of attention over last decades. On one hand, such systems offer a possibility to conveniently test such fundamental concepts as, e.g., macroscopic quantum tunneling \cite{CL} and macroscopic quantum coherence \cite{Leg} as well as to investigate a wealth of novel fascinating transport phenomena. On the other hand, Josephson junction circuits represent a promising platform for a variety of applications, including, e.g., superconducting qubits for quantum information processing \cite{qubits,CQED} and building the quantum standard of electric current \cite{curst,Bloch,ptb1}.

Of particular interest are non-local quantum effects emerging in systems of Josephson junctions coupled through mutual capacitance \cite{SLZ25}. These superconducting structures allow to conveniently explore an interplay between long-range quantum coherence and non-equilibrium dynamics extending the concept of correlated tunneling (or cotunneling) of single electrons \cite{AN90,AN92,GZ92,GAM90} and Cooper pairs \cite{Frank,ptb2,Jukka} to such physical objects as, e.g., magnetic flux quanta (fluxons) which may also tunnel coherently through Josephson nanostructures \cite{LSZ22}.

Recently it was demonstrated \cite{SLZ25} that cotunneling of magnetic fluxons across a pair of capacitively coupled connected in parallel  Josephson junctions gives rise to a novel quantum Coulomb drag effect which amounts to  control the quantum state of one of these junctions by applying an external bias to another one. Note that drag effects in Josephson structures were discussed previously by several research groups \cite{Sakhi,Shi,Cole}. In particular, the authors \cite{Sakhi} investigated drag of superconducting vortices in Josephson junction arrays, while current drag in bilinear Josephson junction chains was studied \cite{Shi,Cole} in the regime of classical Coulomb blockade. In comparison to these and some other earlier works, the predicted quantum Coulomb drag effect  \cite{SLZ25} differs dramatically as it essentially requires establishing macroscopic quantum coherence in the system under consideration.

In this work we will continue investigating a system composed of two coupled Josephson junctions, where a current flowing through one junction may induce a non-vanishing voltage across the other \cite{SLZ25}. The coupling arises solely via the mutual capacitance, otherwise the two junctions would remain electrically isolated from each other. This configuration conveniently allows for a controlled study of coherent tunneling effects which amounts to correlated quantum phase slips occuring (almost) simultaneously in both junctions. While at sufficiently large values of the mutual capacitance these quantum phase slips effectively describe a single tunneling event \cite{SLZ25}, at smaller values of this capacitance this is not anymore so as the correlation between quantum phase slips at different junctions becomes weaker. This more complicated situation requires a special treatment that will be carried out in our present work.

Our primary focus here is to evaluate the fluxon cotunneling amplitudes in the limit of small mutual capacitance which enables one to proceed perturbatively in the interaction. Our results extend the description of quantum Coulomb drag in systems of capacitively coupled superconducting junctions \cite{SLZ25} and provide further insights into a subtle interplay between quantum coherence and interaction effects, offering deeper understanding of quantum transport phenomena in Josephson setups.

The structure of our paper is as follows. In Sec. II we describe our model and outline the general picture of quantum Coulomb drag effect in a system of capacitively coupled Josephson junctions illustrating a direct relation of this effect to cotunneling of fluxons across such junctions. Sec. III is devoted to a perturbative in the interaction analysis of fluxon cotunneling process. Our final results both for cotunneling amplitudes and rates are specified in Sec. IV. Discussion of our main results and observations is presented in Sec. V.

\begin{figure}[h]
   \centering
   \includegraphics[width=8cm]{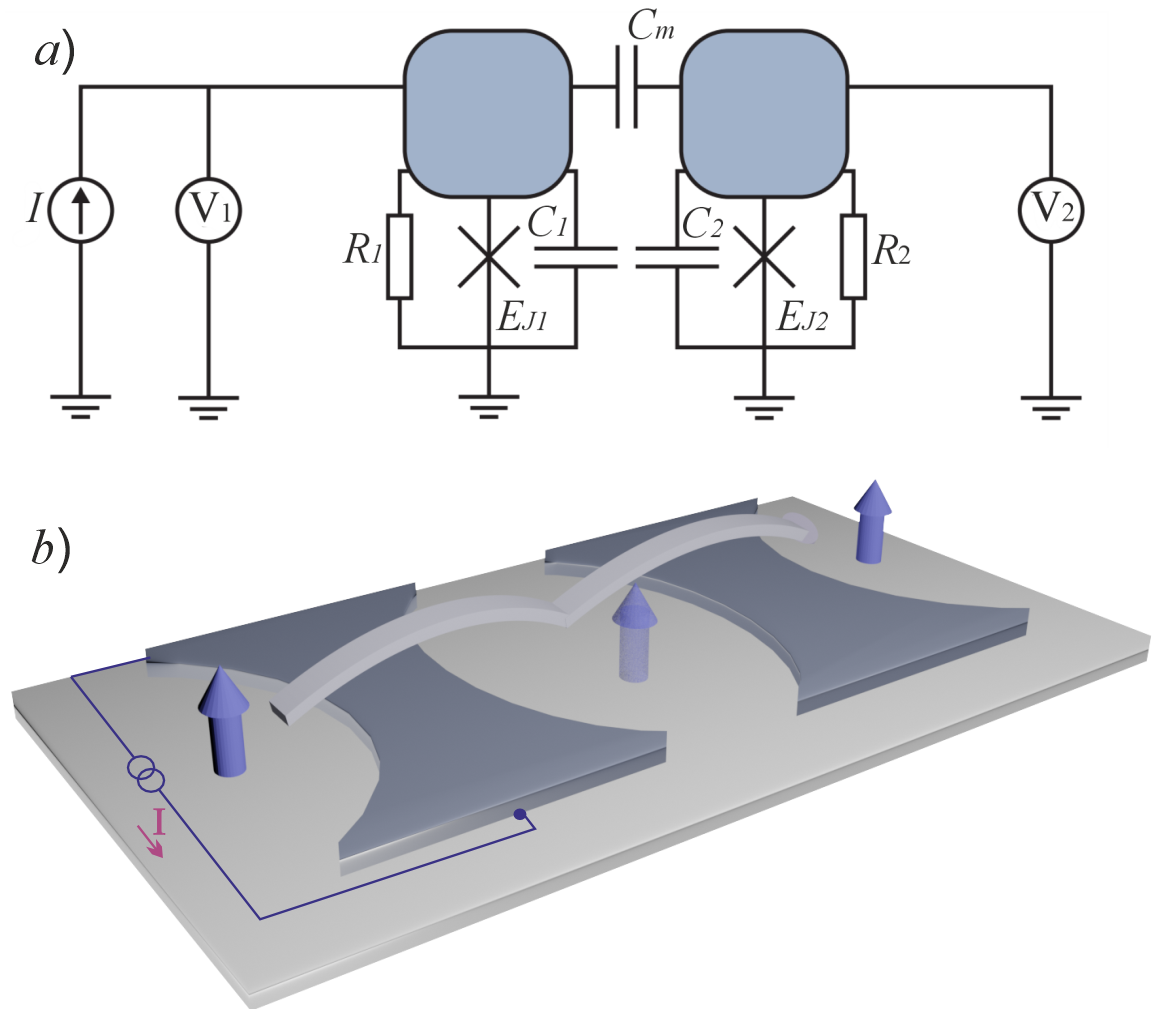}\\
   {Fig. 1: a) The system under consideration. b)  Schematic representation of fluxon cotunneling in a system of coupled Josephson junctions.
.}
\end{figure}

\section{Quantum Coulomb drag in coupled Josephson junctions}
The system under consideration consists of two Josephson junctions coupled via a capacitor $C_{m}$, as it is illustrated in Fig. 1. A small constant current bias $I$ is applied to the left junction whereas no electric current flows through the right one. The grand partition function for this system expressed in terms of the double path integral over the Josephson phase variables at both junctions $\varphi_1$ and $\varphi_2$
\begin{equation}
{\mathcal Z}=\int {\mathcal D}\varphi_1 \int {\mathcal D} \varphi_2 e^{-S[\varphi_1,\varphi_2]}.
\label{pathint}
\end{equation}

The effective action of the system is defined as
\begin{eqnarray}
\nonumber
S[\varphi_1,\varphi_2] &=& \int dt \Big(
\frac{C_1 \dot{\varphi}_1^2}{8e^2} 
+ \frac{C_2 \dot{\varphi}_2^2}{8e^2} 
+ \frac{C_m (\dot{\varphi}_1 - \dot{\varphi}_2)^2}{8e^2} \\
&&+ U(\varphi_1,\varphi_2) 
+ \frac{I \varphi_1}{2e}
\Big) + S_{\text{diss}}, 
\label{action}
\end{eqnarray}
where $C_1$ and $C_2$ are the junction capacitances, the potential energy
\begin{equation}
U(\varphi_1,\varphi_2)=E_{J_{1}} (1 - \cos\varphi_1)+E_{J_{2}} (1 - \cos\varphi_2)
\label{U}
\end{equation}
is formed by their Josephson coupling energies and the term $S_{\text{diss}}$ accounts for linear Ohmic dissipation. It reads \cite{CL,SZ90}:
\begin{equation}
\label{action12} S_\text{diss}=\sum_{k=1,2}\frac{T^2R_Q}{8R_k}\int_{-1/2T}^{1/2T} d\tau \int_{-1/2T}^{1/2T}d\tau'\frac{[\varphi_k (\tau)-\varphi_k(\tau')]^2}{\sin^2 [\pi T(\tau-\tau')]}, 
\end{equation}  
where $R_Q=\pi /(2e^2) \simeq 6.45$ $k\Omega$ is the "superconducting" quantum resistance unit. In what follows we will assume both Ohmic resistances $R_{1,2}$ to obey the condition $R_{1,2}<R_Q$ implying that both Josephson junctions strictly remain on the "superconducting" side of the Schmid phase diagram \cite{Schmid,Blg}. We will also assume that both Josephson coupling energies $E_{J1}$ and $E_{J2}$ strongly exceed the charging ones, i.e. $E_{J1,2} \gg E_{C_{1,2}}=e^2/(2C_{1,2})$.

The action (\ref{action})-(\ref{action12}) describes dissipative dynamics of a quantum "Josephson particle" with "coordinates" $ (\varphi_1, \varphi_2)$ in a two-dimensional (2D) periodic potential $U(\varphi_1, \varphi_2)$ tilted by the external current $I$ along the "axis" $\varphi_1$. Provided this current is small, $I \ll 2eE_{J_{1}}$, and the temperature $T$ is low this quantum particle resides in the vicinity of the potential minima located at $(\varphi_1, \varphi_2) = (2\pi n_1, 2\pi n_2) $ (with integer $n_{1,2}$) and slowly slides down in energy due to quantum tunneling between the minima of  $U(\varphi_1, \varphi_2)$. Each such tunneling event between the neighboring minima corresponds to a quantum phase slip by $\pm 2\pi$ in one of the two junctions and tunneling of one magnetic flux quantum $ \Phi_0 = \pi/e $ across this junction as it is illustrated in Fig. 1b. According to the Josephson relation, this tunneling process results in a voltage drop across the corresponding junction which may generally be expressed in the form  
\begin{eqnarray}\label{volt}
\langle V_k \rangle = \Phi_0 \sum_{n_1, n_2} n_k \Gamma_{n_1, n_2}(I), \quad k=1,2,
\end{eqnarray}
where $ \Gamma_{n_1, n_2}(I) $ denote the tunneling rates between the system states $(0,0)$ and $(2\pi n_1,2\pi n_2)$.

Note that in the limit $I \to 0$ these states remain degenerate. In the absence of dissipation one can define and evaluate quantum mechanical transition amplitudes $\gamma_{n_1,n_2}$  corresponding to the tunneling processes $(0,0) \to (2\pi n_1,2\pi n_2)$. These amplitudes, in turn, can be related to the ground state energy $\varepsilon_0$ which depends on the quasicharge \cite{SZ90} values for our problem  $q_1$ and $q_2$. In the tight binding limit considered here and for the lowest 2D Brillouin zone with $-e<q_1,q_2<e$ one has
\begin{equation}
   \varepsilon_0(q_1,q_2)=-\sum_{n_1,n_2} \gamma_{n_1,n_2}e^{\pi i (n_1q_1+n_2q_2)/e}.
   \label{spectra}
\end{equation} 
Turning on the external current bias $I$ one lifts the degeneracy between the states $(0,0)$ and $(2\pi n_1,2\pi n_2)$. Taking linear Ohmic dissipation into account one readily recovers the relation between the tunneling (decay) rates $\Gamma_{n_1,n_2}(I)$ and the transition amplitudes $\gamma_{n_1,n_2}$. This task can be conveniently accomplished, e.g., by means of the standard trick employing the imaginary part of the free energy \cite{book,Weiss} with the result
\begin{eqnarray}
 \nonumber
 \Gamma_{n_1,n_2}(I)&=&|\gamma_{n_1,n_2}|^2\left(\frac{2\pi T}{\Omega_c}\right)^{2\alpha_{n_1,n_2}}e^{\frac{\pi I n_1}{2eT}}\\
 && \times\frac{\left|\mathit{\Gamma}\left(\alpha_{n_1,n_2}+i\frac{In_1}{2eT}\right)\right|^2}{2\pi T\mathit{\Gamma}(2\alpha_{n_1,n_2})},
\label{Gamma}
\end{eqnarray}
where $\alpha_{n_1,n_2}=R_Q(n_1^2/R_1+n_2^2/R_2)$, $\mathit{\Gamma}(z)$ is the Euler Gamma-function and $\Omega_c$ is a typical frequency of small fluctuations near the potential minima. 

In what follows we will only account for tunneling processes between adjacent potential wells (i.e. we set $n_{1,2}=\pm 1$) since all other processes give a negligible contribution and can be safely neglected.  In the lowest order in the tunneling rates, from Eq. (\ref{volt}) we obtain
\begin{eqnarray}
\label{volt1}
\langle V_{1} \rangle= \Phi_{0}(\Gamma_{1,0}(I)-\Gamma_{-1,0}(I)), \\
\langle V_{2} \rangle= \Phi_{0}(\Gamma_{0,1}(I)-\Gamma_{0,-1}(I)).
\label{volt2}
\end{eqnarray}
Each of these equations defines the voltage drop in one of the junction that occurs due to quantum phase slips in the same junction. In the case of the first junction detailed balance yields 
\begin{equation}
\Gamma_{-1,n}(I)=\exp \left(-I\Phi_0/T\right)\Gamma_{1,n}(I), \quad n=0
\label{dbr}
\end{equation}
and, hence, $\langle V_{1} \rangle$ differs from zero for any non-zero value of $I$ just like in a single current biased Josephson junction \cite{SZ90,book}. The same detailed balance condition for the second junction reads $\Gamma_{0,1}=\Gamma_{0,-1} $ because no current flows across this junction. It follows immediately from Eq. (\ref{volt2}) that $\langle V_{2} \rangle =0$ in the lowest order in tunneling.

Let us now proceed to the next order perturbation theory and take into account the transition rates $\Gamma_{\pm 1, \pm 1}$ corresponding to fluxon tunneling events which occur in {\it both} junctions. In this order, from Eq. (\ref{volt}) we obtain
\begin{eqnarray}
\nonumber
 \langle V_{2} \rangle&=&\Phi_{0}(\Gamma_{1,1}(I)-\Gamma_{1,-1}(I)-\Gamma_{-1,-1}(I)+\Gamma_{-1,1}(I))\\
 &=& \Phi_{0}(\Gamma_{1,1}(I)-\Gamma_{1,-1}(I))(1- \exp \left(-I\Phi_0/T\right)),
 \label{V2} \end{eqnarray}
where we again made use of the detailed balance condition (\ref{dbr}) with $n=\pm 1$. 

One can demonstrate that for $C_m >0$ the combination in Eq. (\ref{V2}) differs from zero, i.e. applying an external current to one Josephson junction one can induce a non-zero average voltage drop $\langle V_{2} \rangle$ across another junction that is capacitively coupled to the first one. This is the essence of the quantum Coulomb drag effect predicted in our recent paper \cite{SLZ25}. This effect is fundamentally linked to quantum phase slips which occur coherently in both junctions and account for the process of fluxon cotunneling \cite{SLZ25,LSZ22}. This process is schematically illustrated in Fig. 1b.

\section{Perturbation theory for fluxon cotunneling}

It follows from the above analysis that in order to describe Coulomb drag effect in the system under consideration it suffices to evaluate the transition amplitudes $\gamma_{1,1}$ and $\gamma_{1,-1}$ for $I=0$ and in the absence of dissipation $1/R_k \to 0$. Then the resulting expressions for $\gamma_{1,\pm 1}$ simply need to be combined with Eqs. \eqref{Gamma} and \eqref{V2} which determine the voltage response $\langle V_{2} \rangle$ across the second junction to the current bias $I$ applied to the first one.

Let us for a moment assume that $C_m=0$. Then the partition function \eqref{pathint} gets factorized into the product of two independent path integrals over the phase variables $\varphi_{1}$ and $\varphi_{2}$, i.e. we have $\mathcal Z=\mathcal Z_1\mathcal Z_2$, where $\mathcal Z_{1,2}$ define the partition functions for each of the two junctions. In the limit $E_{J_k}\gg E_{C_k}$ considered here these partition functions can be evaluated within a semiclassical analysis employing the standard instanton technique. The main contribution to $\mathcal Z_{1,2}$ is determined by fluctuations around multi-instanton trajectories 
\begin{equation}
  \tilde \varphi_{1,2}(\tau)=4\sum_{j=1}^M\nu_j \arctan(e^{\Omega_{1,2}(\tau-\tau_j)}),  
\label{inst} 
\end{equation}
where we define the plasma oscillation frequencies for both junctions $\Omega_k=\sqrt{8 E_{J_{k}}E_{C_{k}}}$, $k=1,2$.  Every $M$-instanton trajectory depends on the topological charges $\nu_j=\pm 1$ and the "positions" $\tau_j$ of individual instantons playing the role of their collective coordinates.  Substituting the trajectories \eqref{inst} into Eq. \eqref{action}, integrating out Gaussian fluctuations around $\tilde \varphi_{1,2}(\tau)$, performing a summation over all "neutral" instanton configurations  $\sum_j\nu_j=0$ and integrating over the instanton coordinates $\tau_j$, we obtain
\begin{equation}
   \mathcal Z_{1,2}=\sum_{M=0}^\infty \frac{\gamma_{1,2}^M}{M!}\int_{-1/2T}^{1/2T} d\tau_1...d\tau_M\sum_{\{\nu_1,..,\nu_M\}}\delta_{0,\sum_j\nu_j},
\label{Z12}
\end{equation}
where 
\begin{equation}
   \gamma_k=4\sqrt{\frac{E_{J_k}\Omega_k}{\pi}}  \exp \left(- \frac{8 E_{J_k}}{\Omega_k}\right), \quad k=1,2.
\end{equation}
The total partition function can be rewritten as
\begin{equation}
   \mathcal Z=\frac{1}{4e^2}\int_{-e}^{e}dq_1\int_{-e}^{e}dq_2 e^{\frac{2\gamma_1\cos(\pi q_1/e)+2\gamma_2\cos(\pi q_2/e)}{T}}.
\label{Z}
\end{equation}
Comparing this expression with the partition function
\begin{equation}
   \mathcal Z \approx \frac{1}{4e^2}\int_{-e}^{e}dq_1\int_{-e}^{e}dq_2 e^{-\frac{\varepsilon_0(q_1,q_2)}{T}}, \quad T\ll \Omega_{1,2},
   \end{equation}
combined with Eq. (\ref{spectra}), one can easily identify
\begin{equation}
   \gamma_{\pm 1,0}=\gamma_1,\qquad \gamma_{0,\pm 1}=\gamma_2.
\end{equation}

Let us now include a non-zero mutual capacitance $C_m$ into our consideration. In this case instantons in different junctions with identical (opposite) topological charges attract (repel) each other \cite{SLZ25}. As a result, for sufficiently large values of $C_{m}$ instantons with identical topological charges merge forming a new instanton that accounts for simultaneous tunneling of both phase variables $\varphi_1$ and $\varphi_2$ between the states (0,0) and $(2\pi,2\pi$) corresponding to the transition amplitude $\gamma_{1,1}$. At the same time, instantons with opposite topological charges cannot merge and, hence, quantum tunneling between the states $(0,0)$ and $(2\pi, -2\pi)$ is only accounted for by two different instantons which provide a much smaller contribution to the corresponding transition amplitude $\gamma_{1,-1}$. In other words, for sufficiently large $C_m$ one always has $\gamma_{1,1} \gg \gamma_{1,-1}$ and, hence, the tunneling process $(0,0) \to (2\pi, -2\pi)$ can be totally neglected \cite{SLZ25}. 

However, at smaller values of $C_{m}$ the interaction between instantons in different junctions becomes weak and they cannot anymore form a single instanton trajectory even for identical topological charges. In order to treat this situation we will proceed perturbatively in $C_{m}$. Substituing the instantons trajectories \eqref{inst} $\tilde{\varphi}_{1}(\tau)$ and $\tilde{\varphi}_{2}(t+\tau)$ (which contain respectively $M_1$ and $M_2$ instantons) into the term proportional to $C_{m}$ in the effective action (\ref{action}), we arrive at the following correction to the action value
\begin{multline}
   \delta S_{m}=\frac{C_m}{8e^2}\int d\tau \left(\frac{d\tilde\varphi_{1}}{d\tau}-\frac{d\tilde\varphi_{2}}{d\tau}\right)^2\\=\frac{M_1\Omega_1+M_2\Omega_2}{2E_{C_m}}+\sum_{j_1=1}^{M_1}\sum_{j_2=1}^{M_2}\nu_{j_1}\epsilon_{j_2}\mathcal V(\tau_{j_1}-\tau'_{j_2}),
   \label{interaction}
\end{multline}
where we defined
\begin{equation}
   \mathcal V(\tau-\tau')=-\frac{\Omega_1\Omega_2}{2E_{C_m}}\int \frac{dt}{\cosh(\Omega_1(t-\tau))\cosh(\Omega_2(t-\tau'))}
\label{inter}
\end{equation}

Note that in the expression for $\tilde{\varphi}_{2}(t+\tau)$ we replaced $\nu_j \to \epsilon_j =\pm 1$ and $\tau_{j} \to \tau_{j}'$ for the sake of clarity. The resulting partition function then reads
\begin{multline}
 \label{partitionf}
    \mathcal Z=\sum_{M_1=0}^\infty\sum_{M_2=0}^\infty \frac{\gamma_1^{M_1}\gamma_2^{M_2}}{M_1!M_1!}\int_{-1/2T}^{1/2T} d\tau_1...d\tau_{M_1}\\
\times\int_{-1/2T}^{1/2T}d\tau_1'...d\tau_{M_2}'\sum_{\{\nu_1,..,\nu_{M_1},\epsilon_1,..,\epsilon_{M_2}\}}\delta_{0,\sum_j\nu_j}\delta_{0,\sum_j\epsilon_j}e^{-\delta S_m}.
  \end{multline}
Combining Eqs. (\ref{interaction}) and  \eqref{partitionf}, we observe that the first term in the right-hand side of Eq. (\ref{interaction}) yields renormalization of the amplitudes $\gamma_k$ as
\begin{equation}
\tilde\gamma_k=\gamma_k \exp \left(-\frac{\Omega_k}{2E_{C_m}}\right), \quad k=1,2,
\label{ren}
\end{equation}
 whereas the remaining combination in the second line of Eq. (\ref{interaction}) defines inter-instanton interactions. 
 
Note that the partition function in Eq. (\ref{partitionf}) formally describes a mixture of two 1D interacting classical gases confined inside a box of the "size" $1/T$. The values $\tilde \gamma_{1,2}$ \eqref{ren} play the role of (small) fugacities for these gases. These gases interact each other and from Eq. (\ref{interaction}) one can observe that the interaction is short-ranged. Thus one can use the well known virial expansion \cite{LLvol5}. In full analogy with the evaluation procedure for the second virial coefficient one can derive the correction to the transition amplitudes caused by the presence of the capacitor $C_m$. For this purpose let us identically rewrite the partition function (\ref{partitionf}) by adding and subtracting its expression in the absence of interaction, i.e. for $\mathcal V(\tau)=0$). Then we obtain
\begin{multline}
   \mathcal Z=\sum_{M_1=0}^\infty\sum_{M_2=0}^\infty \frac{\tilde \gamma_{1}^{M_1}\tilde \gamma_{2}^{M_2}}{M_1!M_2!}\int_{-1/2T}^{1/2T} d\tau_1...d\tau_{M_1}\\
  \times \int_{-1/2T}^{1/2T}d\tau_1'...d\tau_{M_2}'\sum_{\{\nu_1,..,\nu_{M_1}\}}\delta_{0,\sum_j\nu_j}\sum_{\{\epsilon_1,..,\epsilon_{M_2}\}}\delta_{0,\sum_j\epsilon_j}\\+
   \sum_{M_1=1}^\infty\sum_{M_2=1}^\infty \frac{\tilde \gamma_{1}^{M_1}\tilde \gamma_{2}^{M_2}}{M_1!M_2!}\int_{-1/2T}^{1/2T} d\tau_1...d\tau_{M_1}\int_{-1/2T}^{1/2T}d\tau_1'...d\tau_{M_2}'\\\times\sum_{\{\nu_1,..,\nu_{M_1}\}}\delta_{0,\sum_j\nu_j}\sum_{\{\epsilon_1,..,\epsilon_{M_2}\}}\delta_{0,\sum_j\epsilon_j}\\
   \times\left(e^{-\sum_{i_1}\sum_{i_2}\nu_{i_1}\epsilon_{i_2}\mathcal V(\tau_{i_1}-\tau_{i_2}')}-1\right).
\label{virial}
\end{multline}
The first term in the right-hand side of the above expression can be factorized and summed up as above (cf. Eqs. \eqref{Z12}, \eqref{Z}). In order to evaluate the remaining contribution in Eq. \eqref{virial} we observe that the corresponding integrand remains nonzero only provided some of the instanton "coordinates" $\tau_{i_1}$ are located sufficiently close to some of the "coordinates" $\tau_{i_2}'$. It is easy to verify that there exist $M_1M_2$ possible configurations of that kind. Due to the permutation symmetry one can select them as $\tau_1$ and $\tau_1'$ and integrate over these variables separately. As a result, one finds
\begin{multline}
   \mathcal Z\approx\frac{1}{4e^2}\int_{-e}^{e}dq_1dq_2 e^{\frac{2}{T}\left(\tilde\gamma_{1}\cos(\pi q_1/e)+\tilde\gamma_{2}\cos(\pi q_2/e)\right)}\{1+\tilde\gamma_{1}\tilde\gamma_{2}\\ \times\sum_{\nu_1,\epsilon_1}e^{\pi i(q_1+q_2)/e}\int_{-1/2T}^{1/2T} d\tau_1d\tau_1'\big(e^{-\nu_1\epsilon_1\mathcal V(\tau_1-\tau_1')}-1\big)\}.
\end{multline}
Expressing the combination in curly brackets in the form of an exponent we immediately recover the transition amplitudes $\gamma_{\pm1,\pm1}$ which differ from zero to inter-instanton interactions. Specifically, for the interesting for us amplitudes $\gamma_{1,\pm1}$ we obtain
\begin{equation}
   \gamma_{1,\pm 1}=\tilde\gamma_1\tilde\gamma_2\int d\tau\left(e^{\mp\mathcal V (\tau)}-1\right).
\label{g11}
\end{equation}

Equation \eqref{g11} represents the central result of our work perturbalively describing fluxon cotunneling amplitudes in a system of a pair of capacitively coupled Josephson junctions. We also note that the same result \eqref{g11} can also be recovered in a more systematic way with the aid of the so-called Mayer diagram technique \cite{SemenovFuture} enabling one to include other higher order corrections. 

\section{Cotunneling amplitudes and rates}
According to Eq. \eqref{V2} the average voltage $\langle V_2\rangle$ generated across the second junction by the external bias current $I$ applied to the first junction is essentially determined by the difference between the cotunneling rates $\Gamma_{1,1}$ and $\Gamma_{1,-1}$ which account for the tunneling processes from the state (0,0) to the states $(2\pi,2\pi)$ and $(2\pi,-2\pi)$ respectively, see also Fig. 2. Making use of Eq. \eqref{Gamma} we obtain
\begin{multline}
 \Gamma_{1,1}-\Gamma_{1,-1}=(|\gamma_{1,1}|^2- |\gamma_{1,-1}|^2)\\
 \times\left(\frac{2\pi T}{\Omega_c}\right)^{2\alpha}e^{\frac{\pi I}{2eT}}
 \frac{\left|\mathit{\Gamma}\left(\alpha+i\frac{I}{2eT}\right)\right|^2}{2\pi T\mathit{\Gamma}(2\alpha)},
\label{Gammapm}
\end{multline}
where $\alpha=R_Q(1/R_1+1/R_2)$. Hence, in order to recover the induced voltage $\langle V_2 \rangle$ it suffices to evaluate the transition amplitudes $\gamma_{1,\pm 1}$ with the aid of Eq. \eqref{g11}.
\begin{figure}[h]
   \centering
   \includegraphics[width=6cm]{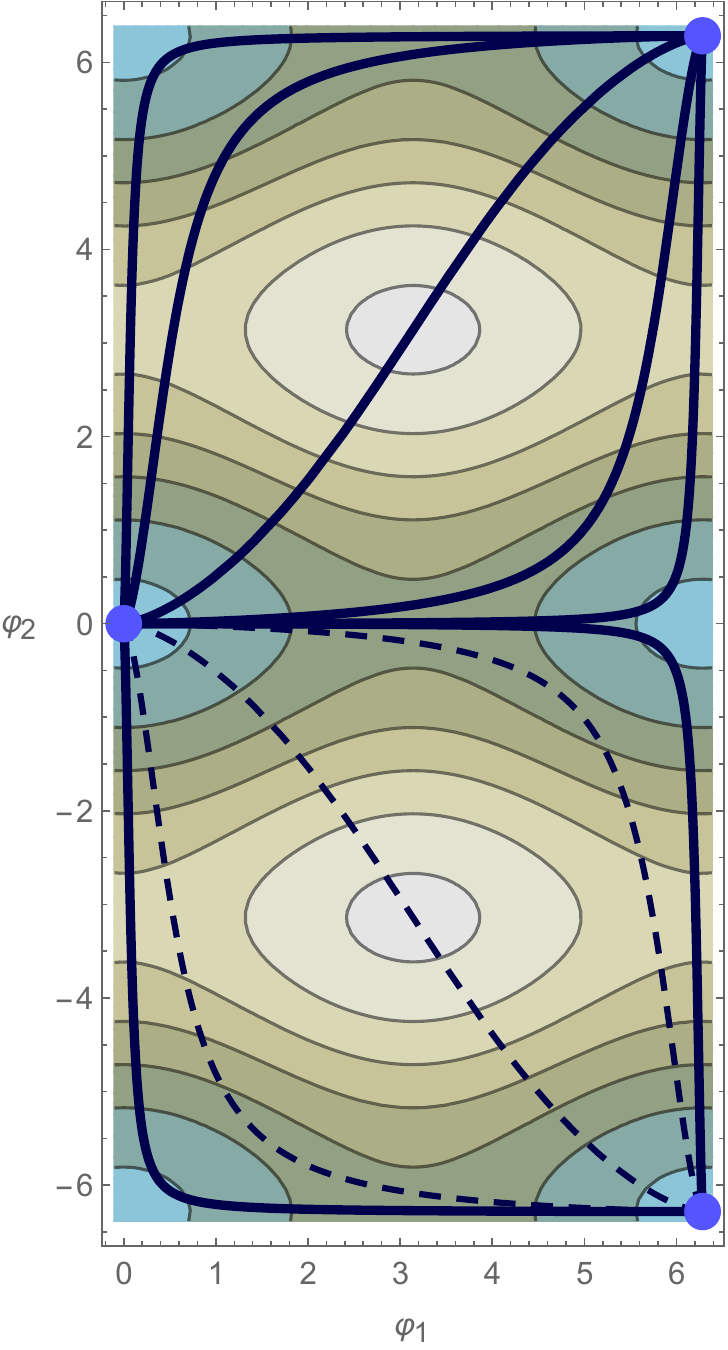}\\
   {Fig. 2:  The saddle point paths in the ($\varphi_1, \varphi_2$) plane controlling the fluxon cotunneling processes with the amplitudes $\gamma_{1,\pm 1}$ for different values of $C_m$. The least action paths are indicated by the solid lines, whereas the saddle point trajectories providing the maximum to the action are depicted by the dashed lines.}
\end{figure}

The qualitative difference between these two amplitudes is illustrated in Fig. 2. For very small values of $C_m$ the interaction between instantons at different junctions is weak and the least action paths (shown by the solid lines) controlling the path integrals for both cotunneling amplitudes remain in the vicinity of the lines ($\varphi_1=0$, $\varphi_2 =\pm 2\pi$) and ($\varphi_2=0$, $\varphi_1 =2\pi$), as displayed in the Figure. With increasing $C_m$ the saddle point paths gradually move towards the centre of the upper ($0<\varphi_{1,2}<2\pi$) and lower ($0<\varphi_{1}, -\varphi_{2} < 2\pi$) squares and eventually merge into the trajectories passing the points $\varphi_{1}=\varphi_{2}=\pi$ and $\varphi_{1}=-\varphi_{2}=\pi$ respectively, see Fig. 2. Since instantons with equal (opposite) topological charges at different junctions attract (repel) each other, these saddle point trajectories in the upper (lower) square of Fig. 2 provide the minimum (maximum) to the action. Accordingly, only trajectories indicated by the solid lines are important, whereas  those shown by the dashed lines are irrelevant and should be disregarded. 

This difference between the two cotunneling processes is, of course, fully captured by Eq. \eqref{g11}. In a symmetric case corresponding to the Josephson junctions with equal parameters $C_{1,2}=C$, $E_{J_{1,2}}=E_{J}$ and $\Omega_{1,2}\equiv\Omega =\sqrt{8 E_{J}/E_{C}}$ this equation yields
\begin{multline}
\gamma_{1,\pm  1} = \frac{16 E_J \Omega}{\pi} e^{-\frac{16 E_J}{\Omega}} e^{- \frac{2 C_m \Omega}{e^2}} \\ \times \int dt   \left( e^{\pm \frac{2 C_m \Omega^2 t}{e^2 \sinh(\Omega t)}} - 1 \right).
\label{gequal}
\end{multline}
In the limit of very small $C_{m} \ll C(\Omega/E_{J})$ it suffices to expand the exponents in Eq. (\ref{gequal}). Then we obtain
\begin{eqnarray}
 \gamma_{1,\pm 1}\simeq \frac{64 E^{2}_{J}\pi}{C \Omega}\left\{\pm C_{m}+\frac{C^{2}_{m}\Omega}{e^{2}}\left(\frac{2}{3}\mp 2\right)\right\}e^{-\frac{16E_{J}}{\Omega}}.
 \label{gsmall}
\end{eqnarray}
Combining Eqs. (\ref{V2}), (\ref{Gammapm}) and (\ref{gsmall}), we get
\begin{multline}
 \langle V_{2}(I) \rangle \simeq  \frac{2^{16}\pi^2}{3}\frac{C^{3}_{m}}{\Omega}   \left(\frac{E^{2}_{J}}{C e}\right)^{2}\sinh\left( \frac{\pi I}{2 e T}\right)e^{-\frac{32E_{J}}{\Omega}}\\
 \times\left(\frac{2\pi T}{\Omega_c}\right)^{2\alpha}
 \frac{\left|\mathit{\Gamma}\left(\alpha+i\frac{I}{2eT}\right)\right|^2}{2\pi T\mathit{\Gamma}(2\alpha)}.
\label{V2pert}
\end{multline}
For larger $C_{m} \gg C(\Omega/E_{J})$ the amplitude $\gamma_{1,-1}$ becomes negligibly small (as it is also illustrated in Fig. 2) and only $\gamma_{1,1}$ matters. The corresponding integral for $\gamma_{1,1}$ in the right-hand side of Eq. (\ref{gequal}) can be handled by means of the steepest descent method. As a result, one finds 
\begin{eqnarray}
 \gamma_{1,1} \simeq 8\sqrt{\frac{3 E_{J}\Omega C}{\pi C_{m}}}e^{-\frac{16E_{J}}{\Omega}}.
 \label{glarge}
\end{eqnarray}
Note that this result exactly coincides with that derived previously by means of a different technique \cite{SLZ25} under the condition $\Omega/E_{J} \ll C_{m}/C \ll 1$. At even larger $C_m \gtrsim C$ the interaction between instantons at different junctions becomes strong and our perturbation theory in $C_m$ ceases to be valid. The latter limit, however, is well described by the semiclassical approach \cite{SLZ25}.

The typical behavior of the cotunneling amplitudes  $\gamma_{1,\pm 1}$ as well as the amplitude $\tilde \gamma \equiv \tilde \gamma_1=\tilde \gamma_2$ (\ref{ren}) depending on the coupling parameter $\lambda = E_{J} / E_{C_{m}}$ is also illustrated in Fig. 3 for a symmetric case of two identical Josephson junctions.
\begin{figure}[h]
   \centering
   \includegraphics[width=9.0cm]{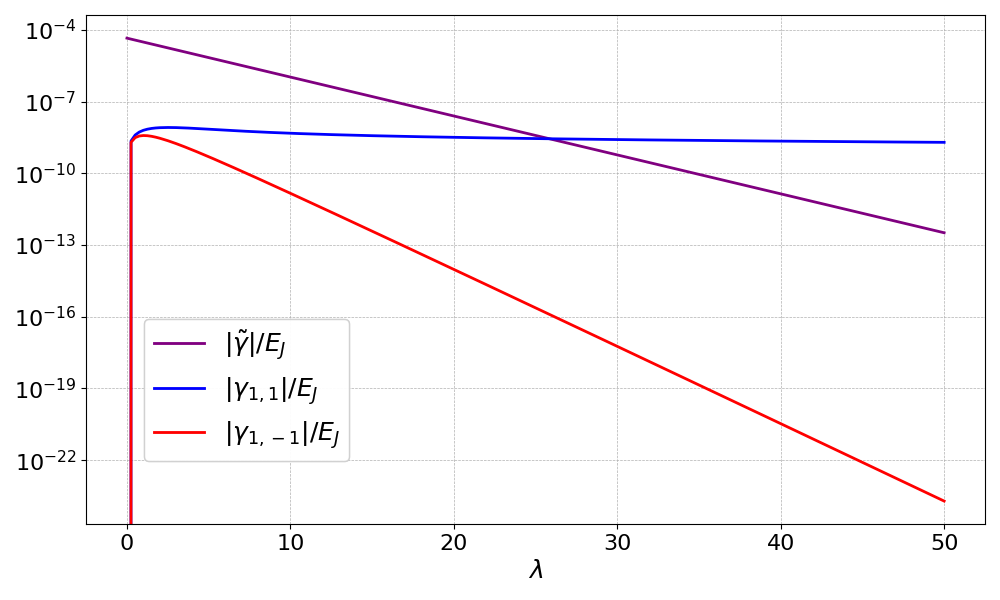}\\
   {Fig. 3: Tunneling amplitudes $\tilde \gamma$ (\ref{ren}) and $\gamma_{1,\pm 1}$ \eqref{gequal} evaluated at $E_C \approx 0.07 E_J$ as functions of the coupling parameter $\lambda = E_{J} / E_{C_{m}}$.}
\end{figure}

One can observe two effects. Firstly, in full accordance with our previous analysis, with increasing the interaction parameter $\lambda$ the cotunneling amplitude $|\gamma_{1,-1}|$ quickly becomes negligibly small as compared to $\gamma_{1,1}$. Secondly, upon further increase of $\lambda$ one eventually enters the regime with $\gamma_{1,1} > \tilde \gamma$ where fluxon cotunneling becomes the dominant process. Remarkably, in this regime not only the induced average voltage $\langle V_2\rangle$, but also $\langle V_1\rangle$ is determined by Eq. \eqref{V2}, i.e. in this case we have $\langle V_1\rangle =\langle V_2\rangle$. We also note that previously the existence of this regime was recognized within a non-perturbative in the interaction analysis \cite{SLZ25}. Here we demonstrated that fluxon cotunneling can dominate over all other tunneling processes also for the parameter range accessible for the perturbation theory.

In order to address the case of different junctions with $E_{C_{1}} \neq E_{C_{2}}$ and  $E_{J_{1}} \neq E_{J_{2}}$  it is 
convenient to introduce the dimensionless parameters $\xi_{k} = \Omega_k / E_{J_{k}}$, $\lambda_{k} = E_{J_{k}} / E_{C_{m}}$ and $\kappa=E_{J_{1}}/E_{J_{2}}$. In terms of these variables, Eq. (\ref{ren}) can be rewritten in the form 
\begin{eqnarray}
 \frac{\tilde{\gamma}_{k}}{E_{J_{k}}}=4\sqrt{\frac{\xi_{k}}{\pi}}e^{-8/\xi_{k}}e^{-\lambda_{k}\xi_{k}/2}, \;\; k=1,2,  
 \label{amnum1}
\end{eqnarray}
whereas the amplitudes $\gamma_{1, \pm 1}$ defined in Eq. (\ref{g11}) read
\begin{multline}
 \frac{\gamma_{1, \pm 1}}{\sqrt{E_{J_{1}}E_{J_{2}}}}=\frac{16}{\pi}\sqrt{\frac{\xi_{1}\kappa}{\xi_{2}}}e^{-8\left(\frac{1}{\xi_{1}}+\frac{1}{\xi_{2}}\right)}e^{-(\lambda_{1}\xi_{1}+\lambda_{2}\xi_{2})/2} \\ \times \int^{+\infty}_{-\infty} dy \left[e^{\pm \frac{\lambda_{1}\xi_{1}}{2} \int  \frac{dx}{\cosh\left((\xi_{1}/\xi_{2}) \kappa x\right)\cosh\left(x-y\right)}} -1\right].
 \label{amnum2}
\end{multline}
The combination in the right-hand side of Eq. \eqref{amnum2} can in general be evaluated only numerically. The corresponding results are displayed in Fig. 4.

\begin{figure}[h]
   \centering
   \includegraphics[width=9.0cm]{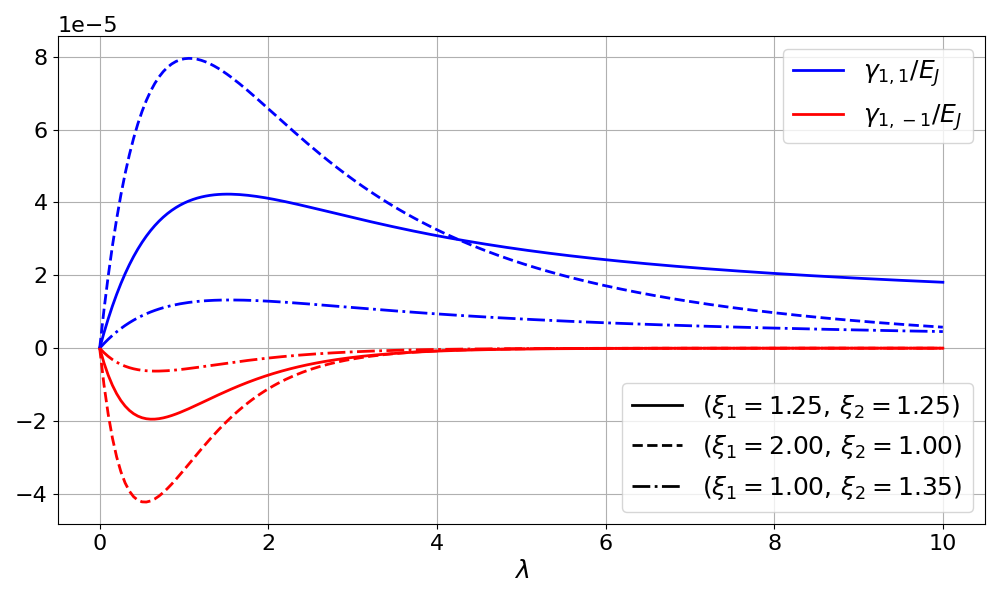}\\
   {Fig. 4: Fluxon cotunneling amplitudes $\gamma_{1,\pm 1}$ \eqref{amnum2} at different values of $\xi_1$ and $\xi_2$. For simplicity we set $\kappa=1$ in which case both amplitudes $\gamma_{1,\pm 1}$ depend on a single interaction parameter $\lambda \equiv \lambda_1=\lambda_2$.}
\end{figure}

We observe that -- similarly to the case of two identical junctions -- the cotunneling amplitude $|\gamma_{1,-1}|$ quickly dies out upon increasing the interaction parameter $\lambda$. Also, one can conclude that for bigger values of $\lambda$ the amplitude $\gamma_{1,1}$ achieves its maximum values in symmetric systems with $\xi_1=\xi_2$.

\section{Discussion}

In this work, we analyzed quantum coherent tunneling dynamics of the Josephson phases in a system of two capacitively coupled resistively shunted superconducting nanojunctions.  Recently it was demonstrated \cite{SLZ25} that such systems may exhibit a novel quantum Coulomb drag effect: By applying an external bias current to one of the junctions one induces a non-zero voltage response not only across this junction but also across the other one that is not biased by any current at all. In a certain sense such "insulating" behavior of the second junction could be taken as a surprise since each of the two junctions separately should remain in the "superconducting" regime with $R_{1,2}<R_Q$.

In the strong coupling limit, i.e. provided the mutual capacitance $C_m$ exceed the junction ones $C$, this effect is associated with "merging" of instantons in the two junctions \cite{SLZ25} corresponding to fluxon cotunneling across the system. On the other hand, at small values of $C_m$ the inter-instanton interaction remains weak and instantons at different junctions cannot anymore merge remaining to a large extent "unbound". Under such conditions the non-perturbative in the interaction approach \cite{SLZ25} becomes obsolete and the very existence of the quantum Coulomb drag in our system could be questioned.

In order to cure this problem, in this work we developed a perturbative in the interaction approach that allows to treat the effects in question for arbitrary small values of $C_m$. Combining the results derived here with those of Ref. \onlinecite{SLZ25} we arrive at the complete description of both fluxon cotunneling and the quantum Coulomb drag effect in the system under consideration. In terms of the effective inter-instanton interaction strength we can distinguish three different regimes of (i) strong coupling $C_m /C \gtrsim 1$, (ii) intermediate coupling $\Omega/E_{J} \lesssim C_{m}/C \lesssim 1$ and (iii) weak coupling  $ 0<C_{m}/C \lesssim \Omega/E_{J}$. While the strong coupling regime (i) needs to be treated non-perturbatively in the interaction \cite{SLZ25}, the weak coupling one (iii) can only be accessed with the aid of the perturbation theory derived here. It is also satisfactory to observe that both these approaches overlap in the intermediate coupling regime (ii) thereby allowing for a smooth and transparent transition between strong and weak coupling limits (i) and (iii).

Our basic conclusion reached here is that the quantum Coulomb drag effect survives at any non-zero value of the mutual capacitance $C_m$.
The specifics of the regime (iii) is that in order to obtain the correct results it is necessary to consider the trade off between two different cotunneling processes $(0,0) \to (2\pi,2\pi)$ and $(0,0) \to (2\pi,-2\pi)$, cf. also Fig. 2. At the same time, only the first of these two processes matters in the regimes (i) and (ii). Perhaps we can also add that despite the parametrical smallness of the non-local voltage response $\langle V_2 \rangle \propto C_m^3$ in the weak coupling regime the actual magnitude of this voltage signal may not be so small due to an unusually large numerical prefactor in Eq. \eqref{V2pert}.

Further increase of the magnitude of the effect can be expected, e.g., in bilinear superconducting granular arrays and Josephson chains. The results derived here as well as in Ref. \onlinecite{SLZ25} can be directly generalized to such systems provided one can neglect the effect of the grain capacitances to the ground. Should these capacitance effects matter the ground state properties of such arrays and chains are known to become richer \cite{BFSZ} and the corresponding modifications to our analysis would be necessary. This issue goes beyond the frames of the present paper and will be addressed elsewhere.

\end{document}